\documentclass[useAMS,usenatbib,usegraphicx]{mn2e}
\usepackage{times}

\newcommand{\kep}{{\em Kepler}}

\newcommand{\kepmi}{{\em Kepler Mission}}
\newcommand{\msol}{\ensuremath{\rm{M}_{\odot}}}

\newcommand{\teff}{\ensuremath{T_{\rm{eff}}}}
\newcommand{\logg}{\ensuremath{\log g}}
\newcommand{\logy}{\ensuremath{\log y}}

\newcommand{\twom}{{\sc 2mass}}
\newcommand{\corr}{\ensuremath{F_{\rm{cont}}}}
\newcommand{\fmin}{\ensuremath{f_{\rm{min}}}}
\newcommand{\fmax}{\ensuremath{f_{\rm{max}}}}
\newcommand{\fmed}{\ensuremath{f_{\rm{med}}}}
\newcommand{\Fmax}{\ensuremath{f_{\rm{+}}}}
\newcommand{\Amax}{\ensuremath{A_{\rm{+}}}}
\newcommand{\tFmax}{\multicolumn{1}{c}{\ensuremath{f_{\rm{+}}}}}
\newcommand{\tAmax}{\multicolumn{1}{c}{\ensuremath{A_{\rm{+}}}}}
\newcommand{\sig}{\ensuremath{\sigma}}
\newcommand{\uHz}{\ensuremath{\umu{\rm{Hz}}}}

\newcommand{\kic}{{KIC}}
\newcommand{\mkep}{\ensuremath{Kp}}
\newcommand{\tabkic}{\multicolumn{1}{c}{\kic}}
\newcommand{\tsig}{\multicolumn{1}{c}{\sig}}
\newcommand{\tpuHz}{\multicolumn{1}{c}{(\ensuremath{\umu{\rm{Hz}})}}}
\newcommand{\tpsig}{\multicolumn{1}{c}{(\sig)}}

\newcommand{\flc}{\ensuremath{f_{\rm{LC}}}}
\newcommand{\fsc}{\ensuremath{f_{\rm{SC}}}}
\newcommand{\fnyq}{\ensuremath{f_{\rm{Nyq}}}}
\newcommand{\papI}{{Paper~{\sc i}}}

\title[First Kepler results on compact pulsators VI]
{First {\em Kepler} results on compact pulsators\\
VI. Targets in the final half of the survey phase}
\author[R.~H.~{\O}stensen et al.]
       {R.~H.~{\O}stensen,$^1$\thanks{E-mail: roy@ster.kuleuven.be}
        R.~Silvotti,$^{2}$                
        S.~Charpinet,$^{3}$                
        R.~Oreiro,$^{1,4}$                     
        S.~Bloemen,$^1$                      
        A.~S.~Baran,$^{5,7}$                 
\newauthor
        M.~D.~Reed,$^{6}$
        S.~D.~Kawaler,$^{7}$                 
        J.~H.~Telting,$^{8}$                 
        E.~M.~Green,$^{9}$                   
        S.~J.~O'Toole,$^{10}$                
        C.~Aerts,$^{1,11}$                
\newauthor
        B.~T.~G\"ansicke,$^{12}$                
        T.~R.~Marsh,$^{12}$                
        E.~Breedt,$^{12}$                
        U.~Heber,$^{13}$                     
        D.~Koester,$^{14}$                   
        A.~C.~Quint,$^{6}$                
\newauthor
        D.~W.~Kurtz,$^{15}$                
        C.~Rodr{\'{\i}}guez-L{\'o}pez,$^{16,17}$                
        M.~Vu\v{c}kovi\'{c},$^{1,18}$                
        T.~A.~Ottosen,$^{8,19}$                
\newauthor
        S.~Frimann,$^{8,19}$                
        A.~Somero,$^{8,20}$                
        P.~A.~Wilson,$^{8,21}$                
        A.~O.~Thygesen,$^{8}$                
        J.~E.~Lindberg,$^{8,22}$                
\newauthor
        H.~Kjeldsen,$^{19}$
        J.~Christensen-Dalsgaard,$^{19}$                
        C.~Allen,$^{23}$
        S.~McCauliff$^{23}$
\newauthor and
        C.~K.~Middour$^{23}$\\
        $^1$Instituut voor Sterrenkunde, K.~U.~Leuven, Celestijnenlaan 200D,
        3001 Leuven, Belgium\\
        $^2$INAF-Osservatorio Astronomico di Torino,
        Strada dell'Osservatorio 20, 10025 Pino Torinese, Italy \\
        $^3$Laboratoire d'Astrophysique de Toulouse-Tarbes, Univ.~de Toulouse,
        14 Av.~Edouard Belin, Toulouse 31400, France \\
        $^4$Instituto de Astrof\'isica de Andaluc\'ia,
        Glorieta de la Astronom\'ia s/n, 18008 Granada, Spain \\
        $^{5}$Mt.~Suhora Observatory, Cracow Pedagogical University,
        Podchorazych 2, 30-084 Krakow, Poland\\
        $^{6}$Department of Physics, Astronomy, and Materials Science,
        Missouri State University, Springfield, MO 65804, USA \\
        $^{7}$Department of Physics and Astronomy, Iowa State University,
        Ames, IA 50011, USA \\
        $^{8}$Nordic Optical Telescope, 38700 Santa Cruz de La Palma, Spain \\
        $^{9}$Steward Observatory, University of Arizona, 933 N.~Cherry Ave.,
        Tucson, AZ 85721, USA \\
        $^{10}$Australian Astronomical Observatory, PO Box 296, Epping NSW 1710,
        Australia\\
        $^{11}$Department of Astrophysics, IMAPP, Radboud University
        Nijmegen, 6500 GL Nijmegen, The Netherlands \\
        $^{12}$Department of Physics, University of Warwick,
        Coventry CV4 7AL, UK \\
        $^{13}$Dr. Karl Remeis-Observatory \& ECAP, Astronomical Inst.,
        FAU Erlangen-Nuremberg, Sternwartstr.~7, 96049 Bamberg, Germany \\
        $^{14}$ Institut f\"ur Theoretische Physik und Astrophysik,
        Universit\"at Kiel, 24098 Kiel, Germany\\
        $^{15}$ Jeremiah Horrocks Institute of Astrophysics,
        University of Central Lancashire, Preston PR1 2HE, UK \\
        $^{16}$Departamento de F\'\i sica Aplicada, Univ. de Vigo,
        Campus Lagoas-Marcosende s/n, 36310 Vigo, Spain \\
        $^{17}$University of Delaware, Department of Physics and Astronomy,
        217 Sharp Lab, Newark, DE 19716, USA\\
        $^{18}$European Southern Observatory, Alonso de C\'ordova 3107,
        Vitacura, Casilla 19001, Santiago, Chile \\
        $^{19}$Department of Physics and Astronomy, Aarhus University,
        8000 Aarhus C, Denmark \\
        $^{20}$Tuorla Observatory, Department of Physics and Astronomy,
        University of Turku, V\"ais\"al\"antie 20, FI-21500, Piikki\"o, Finland\\
        $^{21}$Astrophysics Group, School of Physics, University of Exeter,
        Stocker Road, Exeter EX4 4QL, UK\\
        $^{22}$Centre for Star and Planet Formation, Nat.~Hist.~Museum of Denmark,
        Univ.~of Copenhagen, {\O}. Voldgade 5-7, Copenhagen DK-1350, Denmark\\
        $^{23}$Orbital Sciences Corporation/NASA Ames Research Center, 
        Moffett Field, CA 94035, USA \\
}
\begin{document}

\date{Released 2010 Xxxxx XX}
\pagerange{\pageref{firstpage}--\pageref{lastpage}} \pubyear{2010}
\maketitle
\label{firstpage}

\begin{abstract}
We present results from the final six months of a survey to search for
pulsations in white dwarfs and hot subdwarf stars with the \kep\ spacecraft.
Spectroscopic observations are used to separate the objects into
accurate classes, and we explore the physical parameters of the
subdwarf~B (sdB) stars and white dwarfs in the sample.
From the \kep\ photometry and our spectroscopic data, we find 
that the sample contains 5 new pulsators of the V1093\,Her type,
one AM\,CVn type cataclysmic variable, and a number of other
binary systems.

This completes the survey for compact pulsators with \kep.
No V361\,Hya type of short-period pulsating sdB stars were
found in this half, leaving us with a total of one single multiperiodic
V361\,Hya and 13 V1093\,Her pulsators for the full survey.
Except for the sdB pulsators, no other clearly pulsating hot subdwarfs or
white dwarfs were found, although a few low-amplitude candidates
still remain. The most interesting targets discovered in this
survey will be observed throughout the remainder of the \kepmi,
providing the most long-term photometric datasets ever made on such
compact, evolved stars. Asteroseismic investigations of these
datasets will be invaluable in revealing the interior structure of
these stars, and will boost our understanding of their evolutionary
history.

\end{abstract}


\section{Introduction}

The \kep\ spacecraft was launched in March 2009, with the primary aim
to find Earth-sized planets within the habitable zone around solar-like stars
using the transit method \citep{borucki10}.
In order to have a high probability of finding such planets, the spacecraft
continuously monitors the brightness of $\sim$100\,000 stars with
close to micromagnitude precision.
As a byproduct of the planet hunt, high quality photometric data of
variable stars are obtained, an incredibly valuable input for
the study of binary stars \citep{prsa10} and asteroseismology
\citep{gilliland10a}.

The first four quarters of the \kepmi\ were dedicated to a survey phase,
and a substantial number of target slots for short cadence observations
were made available to the \kep\ {\em Asteroseismic Science Consortium}
({\sc kasc}); 512 slots in the initial roll position (Q1) and 140
in the following quarters.
The series of papers of which this is the sixth,
deals with the search for compact pulsators
\footnote{
The term `compact pulsators' is used to encompass all the various groups
of pulsating white dwarfs and hot subdwarf stars.
}, and the 
results from the first half of the survey are described in
\papI~\citep{KepI}. Paper~{\sc ii} \citep{kawaler10a} describes \kic\,10139564,
a short-period subdwarf~B pulsator (V361\,Hya star).
Five long-period sdB pulsators
(V1093\,Her stars) are described in Paper~{\sc iii} \citep{reed10a}, and one
of them, KPD\,1943+4058, is given a detailed asteroseismic analysis in
Paper~{\sc iv} \citep{VanGrootel10}. An asteroseismic analysis on another 
of the stars in Paper~{\sc iii}, \kic\,2697388, is given by \citet{charpinet10}.
Two more V1093\,Her pulsators that appear to be in
short-period binary systems with M-dwarf companions are described in
Paper~{\sc v} \citep{kawaler10b}. 

The first half of the survey also revealed that the
eclipsing sdB+dM binary, 2M1938+4603 (\kic\,9472174), is a low amplitude
pulsator with an exceptionally rich frequency spectrum \citep{2m1938}, and the
sdB+WD binary, KPD\,1946+4340 (\kic\,7975824),
was found to show eclipses, ellipsoidal modulation and Doppler
beaming effects \citep{bloemen10}.
The current paper describes the content of the compact pulsator sample
observed in the second half of the survey phase.

The \kep\ field of view covers 105 square degrees, and is being observed
in a broad bandpass (4200--9000\,\AA) using 42 CCDs mounted in pairs
on 21 modules.
Although the \kep\ photometer samples the field every 6.54\,s, telemetry
restrictions do not permit the imaging data to be downloaded.
Instead, pixel masks of targets deemed to be of interest must be uploaded
to the spacecraft, and these pixels are averaged into samples of either
approximately one minute (short cadence; SC)
or half an hour (long cadence; LC).

The primary goal of the asteroseismology survey phase is therefore to
identify the most
interesting pulsators in the sample, so that these objects can then be
followed throughout the remaining years of the mission.
The primary goals for the compact pulsator survey were set out
in \papI, and for the subdwarf B pulsators a substantial list
of targets for further study were identified. However, no clearly
pulsating white dwarf was found in the first part of the sample,
and it was hoped that the second part would bring more luck.
This did not happen, so we are still without any confirmed white dwarf
pulsators to follow for the specific target part of the \kepmi.
Analysis of the five unambiguous V1093\,Her pulsators found in
this second half of the survey are
presented by \citet[][Paper~{\sc vii}]{baran10c}.
A study of the period spacings observed in many of the V1093\,Her
stars from the survey has been given by \citet[][Paper~{\sc viii}]{reed10c}.

For an introduction to the pulsating subdwarf\,B stars, we refer the
reader to the earlier papers in this series.
In the present article we will provide photometric variability limits on all
the stars from the second half of the sample and physical parameters for
the hot subdwarf stars, as we did in \papI. Moreover, we will also provide
physical data from our spectroscopy on the white dwarf stars.

We will also present analysis of the \kep\ photometry for a number of
objects that display long-period variability features, and for many of these
we conclude that they are most likely to be binary systems composed of
a hot subdwarf and a white dwarf or main sequence star.

\begin{table*}
\caption[]{Compact pulsator candidates observed with \kep~in Q3 and Q4.}
\label{tbl:targets}
\centering
\begin{tabular}{rllllllll} \hline
\tabkic  & Name          & Run  & RA(J2000) & Dec(J2000) &
          \multicolumn{1}{c}{\mkep}&\corr& Sample & Class \\ \hline
 2020175&J19308+3728&Q3.1&19:30:48.5&+37:28:19 & 15.49 &0.722& ce$^\dag$ & sdB \\
 2303576&J19263+3738&Q3.3&19:26:18.9&+37:38:15 & 17.45 &0.928& c & He-sdO \\
 2304943&J19275+3738&Q3.3&19:27:33.8&+37:38:55 & 16.18 &0.692& c & sdB \\
 2850093&J19237+3801&Q3.2&19:23:47.2&+38:01:44 & 14.73 &0.298& f & F \\
 3343613&J19272+3827&Q3.2&19:27:15.0&+38:27:19 & 15.74 &0.469& df& He-sdOB \\
 3353239&J19367+3825&Q4.1&19:36:46.3&+38:25:27 & 15.15 &0.099& f & sdB \\
 3527028&J19024+3840&Q4.2&19:02:25.7&+38:40:20 & 17.09 &0.465& c & sdB \\
 3938195&J19048+3903&Q4.1&19:04:49.5&+39:03:16 & 15.30 &0.908& f & F \\
 4547333&J19082+3940&Q3.3&19:08:17.1&+39:40:36 & 16.32 &0.253& c$\ddag$ & AM\,CVn\\
 5340370&J18535+4035&Q4.2&18:53:31.1&+40:35:19 & 17.08 &0.128& c & sdB \\
 5557961&J19514+4043&Q4.3&19:51:26.2&+40:43:36 & 15.82 &0.647& f & F \\
 5769827&J18547+4105&Q4.1&18:54:45.0&+41:05:15 & 16.62 &0.952& c & DA0 \\
 5938349&J18521+4115&Q3.2&18:52:10.1&+41:15:15 & 16.05 &0.079& c & sdB \\
 6371916&J19370+4145&Q3.3&19:37:01.1&+41:45:39 & 14.97 &0.361& e & B \\
 6522967&J19279+4159&Q3.2&19:27:58.7&+41:59:03 & 16.91 &0.622& d & sdB \\
 6614501&J19368+4201&Q3.3&19:36:50.0&+42:01:44 & 16.09 &0.600& f & sdB \\
 6878288&J19436+4220&Q3.1&19:43:37.0&+42:20:58 & 16.67 &0.686& f & He-sdOB \\
 7104168&FBS1907+425&Q3.1&19:08:45.7&+42:38:32 & 15.48 &0.189& af& sdB \\
 7129927&J19409+4240&Q3.1&19:40:59.4&+42:40:31 & 16.59 &0.585& e$^\dag$ & DA+DA \\
 7335517&J18431+4259&Q3.2&18:43:06.7&+42:59:18 & 15.75 &0.295& df& sdO+dM \\
 7668647&FBS1903+432&Q3.1&19:05:06.2&+43:18:31 & 15.40 &0.226& af& sdBV \\
 7799884&J18456+4335&Q4.1&18:45:37.2&+43:35:25 & 16.87 &0.109& d & sdB \\
 8054179&J19569+4350&Q3.1&19:56:55.6&+43:50:17 & 14.43 &0.093& f & He-sdOB \\
 8302197&J19310+4413&Q3.3&19:31:03.4&+44:13:26 & 16.43 &0.256& f & sdBV \\
 8874184&J19084+4508&Q4.1&19:08:24.7&+45:08:32 & 16.52 &0.091& d & sdB \\
 9095594&J19369+4526&Q3.2&19:36:59.4&+45:26:27 & 17.69 &0.434& d & sdB \\
 9211123&J19144+4539&Q3.3&19:14:27.7&+45:39:10 & 16.10 &0.447& d & sdB \\
 9637292&J19030+4619&Q3.1&19:03:02.0&+46:19:55 & 16.68 &0.513& f & B \\
10001893&J19095+4659&Q3.2&19:09:33.5&+46:59:04 & 15.85 &0.710& df& sdBV \\
10149211&J19393+4708&Q4.2&19:39:18.3&+47:08:55 & 15.52 &0.240& f & sdB \\
10198116&J19099+4717&Q4.1&19:09:59.4&+47:17:10 & 16.41 &0.238& de& DA \\
10207025&J19260+4716&Q3.3&19:26:05.9&+47:16:31 & 15.04 &0.068& f & He-sdO \\
10449976&J18472+4741&Q3.2&18:47:14.1&+47:41:47 & 14.86 &0.006& df& He-sdOB \\
10462707&J19144+4737&Q4.1&19:14:29.1&+47:37:41 & 16.89 &0.072& d & sdB \\
10553698&J19531+4743&Q4.1&19:53:08.4&+47:43:00 & 15.13 &0.385& f & sdB \\
10579536&J18465+4751&Q3.1&18:46:33.9&+47:51:08 & 17.10 &0.800& e$^\dag$ & B \\
10784623&J19045+4810&Q4.2&19:04:34.9&+48:10:22 & 16.95 &0.065& d & sdB \\
10789011&J19136+4808&Q3.2&19:13:36.3&+48:08:24 & 15.50 &0.031& df& sdOB \\
10961070&J18534+4827&Q4.2&18:53:29.5&+48:27:52 & 16.99 &0.970& d & sdOB \\
10966623&2M1908+4829&Q3.2&19:08:12.8&+48:29:35 & 14.87 &0.030& bf& B \\
11337598&J18577+4909&Q3.3&18:57:47.3&+49:09:38 & 16.11 &0.225& d & DA \\
11350152&J19268+4908&Q3.1&19:26:51.5&+49:08:49 & 15.49 &0.023& d & sdB+F/G \\
11400959&J19232+4917&Q4.1&19:23:17.2&+49:17:31 & 16.89 &0.577& d & sdB \\
11558725&J19265+4930&Q3.3&19:26:34.1&+49:30:30 & 14.95 &0.028& f & sdBV \\
11604781&J19141+4936&Q3.1&19:14:09.0&+49:36:41 & 16.72 &0.006& e$^\dag$ & DA \\
12021724&J19442+5029&Q4.2&19:44:12.7&+50:29:39 & 15.59 &0.558& e$^\dag$ & sdB \\
12069500&J19419+5031&Q4.1&19:41:58.6&+50:31:09 & 13.63 &0.654& f & F \\ \hline
\end{tabular}\\
{\em Notes.}---The decimal point to the run numbers indicates the relevant
month of the mission quarter.\\
\mkep\ is the magnitude in the \kep\ bandpass.
\corr\ is the contamination factor from the \kic~(zero is no
contamination). \\ The samples are:
a: Literature, b: \twom, c: SDSS, d: {\sc galex},
e: Reduced proper motion (RPM), f: \kic\ color.\\ 
A $^\dag$ marks targets with TNG photometry.
$\ddag$: Described in \citet{fontaine10}.
\end{table*}

\section{Survey sample}\label{sect:select}

The methods used to select the sample stars were described in detail in
\papI. In brief, three groups submitted targets based on six different
selection methods, which we designate a -- f in Table~\ref{tbl:targets}.
Only two stars in the current half of the sample were already classified
as compact stars from earlier surveys (sample a),
FBS\,1907+425 and FBS\,1903+432 \citep{abrahamian90}.
The first one of these also appears in the \twom\ color-selected sample (b),
but was dropped as a candidate after follow-up spectroscopy demonstrated
that the star was a normal B-star.  It did, however, reenter the survey
sample through the \kic\ color selected sample (f).
Eight stars appear in the {\sc segue} extension \citep{SEGUE}
of the Sloan Digital Sky Survey \citep[SDSS,][]{SDSS} (c), and all these
are compact objects.
Seventeen of the stars were selected based on UV-excess from {\em Galex}
satellite data \citep[][]{GALEX} (d),
and these are also all hot subdwarfs or white dwarfs.
Seven stars were chosen based on their position in the reduced proper
motion (RPM) diagram (e), but only five of these turned out to be
compact objects. 
As many as 22 stars were selected based only on \kic\ $gri$ colors (f),
and of these, 14 were not included in any of the other samples. Of the
latter, 9 were compact stars and five normal B -- F stars.
In total, four B- and four F-stars contaminate our sample of 47 targets,
and will not be discussed further here.
The remaining compact pulsator candidates have been spectroscopically
classified as WDs or WD composites (6 objects), 
and hot subdwarfs or subdwarf composites (33 objects), as described
in the following section.

One target in the final sample, \kic\,10784623, was scheduled for observations
but was not observed, as it happens to be located on module
3 which suffered a technical fault early in the fourth quarter
(see Section~\ref{sect:instr}), and is no longer in use.
Since the spacecraft is rotated every quarter in order to keep its sunshade
facing in the right direction, it could be possible to observe it at some
time later in the mission.

\begin{table}
\caption[]{Log of spectroscopic observations.}
\label{tbl:spectro}
\centering
\begin{tabular}{lllrl} \hline
Run& Dates                 & Telescope & P.I., Observer \\ \hline
N1 & 2008 September 20--21 & NOT & JHT, AS \\
N2 & 2008 September 22--26 & NOT & RO \\
W1 & 2009 April 11--12     & WHT & CA, RH{\O} \\
W2 & 2009 July 14--16      & WHT & CA, TAO \\
N3 & 2009 September 7      & NOT & JHT, JL \\
N4 & 2010 June 9           & NOT & JHT, AT \\
W3 & 2010 July 2--6        & WHT & CA, RH{\O} \\
N5 & 2010 September 27     & NOT & JHT, SF, PW \\
\hline
\end{tabular}
\end{table}

\section{Spectroscopy}

All the targets in
Table~\ref{tbl:targets} were observed with low resolution spectrographs at
various telescopes, as listed in Table~\ref{tbl:spectro}.
The observations at the {\em Nordic Optical Telescope} ({\sc not}) were done
with the {\sc alfosc} spectrograph, with
grism \#6 in 2008, and grism \#14 in after that.
Both give $R$\,$\approx$\,600
for the $\sim1''$ slit we used, and $\lambda$\,=\,3300\,--\,6200\,\AA.
On the {\em William Herschel Telescope} (WHT)
we used the {\sc isis} spectrograph with grating R300B on
the blue arm (R\,$\approx$\,1600, $\lambda$\,=\,3100\,--\,5300\,\AA).
Red arm spectra were also obtained, but have not been used for this
work.
All data were reduced with the standard {\sc iraf} procedures for
long-slit spectra.

\begin{figure}
\centering
\includegraphics[width=\hsize]{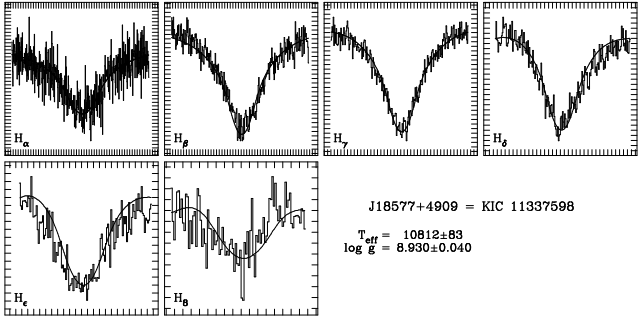}
\caption{
Spectrum and line-profile fit to \kic\,11337598, assuming a rotational
broadening of 1500\,km/s. The spectrum, in particular the higher-resolution
red-arm part, is rather noisy, but the extraordinary broad cores are
clearly evident.
The ticks mark 5\,\AA\ on the X-axis, and 2 percent of the continuum
level on the Y-axis.
The uncertainties stated on the figure are the formal fitting errors.
The real errors are much larger since the RV and $v$sin$i$ are not
fitted simultaneously.
}
\label{fig:K11337598}
\end{figure}

\subsection{The white dwarfs}\label{sect:wds}

There are only five white dwarfs contained in the second half of
the survey sample, and all show Balmer line dominated spectra typical
of DA white dwarfs. 
One object was initially classified as a DB star in our survey, and
caused quite some excitement when the \kep\ light curve was released
displaying clear variability with periods resembling a V777\,Her pulsator.
However, after several efforts to fit the broad helium lines
with DB model atmospheres failed, it was realised that the
object is actually an AM\,CVn type of cataclysmic variable, in which
helium is accreted onto a white dwarf.
This object, \kic\,4547333, cannot be analysed with the
model spectra used for this paper. A suitable grid of model spectra
developed especially for the analysis of this interesting system
is described in \citet{fontaine10}. Since all
details of \kic\,4547333 are provided in that paper,
we will not discuss it further here.

In Table~\ref{tbl:wds} we list the five WDs of the current sample together
with the seven DAs and the DB from \papI.
We have attempted to fit model grid spectra to these WDs, using the same
procedure as for the sdBs and the grids described in \citet{koester10}.

\begin{figure}
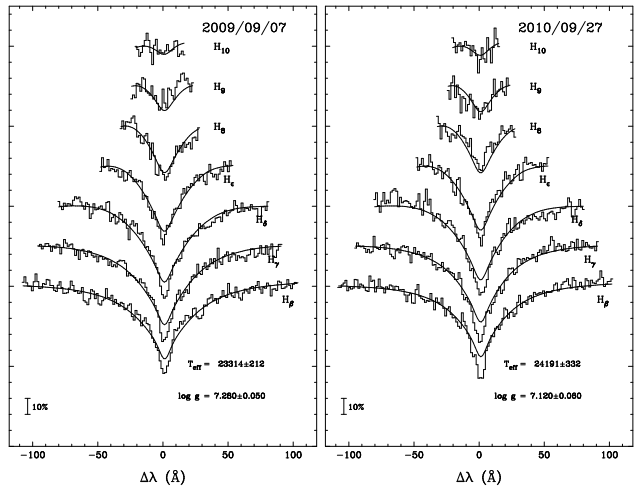

\centering
\includegraphics[width=0.49\hsize]{J19409+4240a.eps}
\includegraphics[width=0.49\hsize]{J19409+4240b.eps}
\caption{
Spectrum and line-profile fit to two observations of \kic\,7129927,
taken a year apart.
Both spectra show that the target has unusually narrow Balmer line cores
superimposed on a normal broad DA profile.
The uncertainties stated on the figure are the formal fitting errors,
but as the fit is obviously far from adequate and the object most
likely composite, the best fit solution can be quite far from
the the true parameters of the brightest system component.
}
\label{fig:K7129927}
\end{figure}

\begin{table}
\caption[]{Spectroscopic properties of the white dwarfs, including the 8 from
\papI\ (marked with \dag).}
\label{tbl:wds}
\tabcolsep 4pt
\begin{tabular}{llccll} \hline
         & Survey       &\teff  &\logg  & & \\
\tabkic  & name         & (kK)  &(dex)  & Class & Run\\ \hline
{\tt\ }3427482$^\dag$ & J19053+3831  &\multicolumn{2}{c}{Poor fit}& DA & W2 \\
{\tt\ }4829241$^\dag$ & J19194+3958  &19.4(5)&7.8(3)& DA2 & W1 \\
{\tt\ }5769827        & J18547+4105  &66(2)  &8.2(3)& DA0 & W1 \\
{\tt\ }6669882$^\dag$ & J18557+4207  &30.5(5)&7.4(3)& DA1 & W2 \\
{\tt\ }7129927        & J19409+4240  &\multicolumn{3}{c}{Composite DA2+DA3}& N3,N5\\
{\tt\ }8682822$^\dag$ & J19173+4452  &23.1(5)&8.5(3)& DA2 & W1 \\
{\tt\ }9139775$^\dag$ & J18577+4532  &24.6(5)&8.6(3)& DA2 & W2 \\
10198116        & J19099+4717  &14.2(5)&7.9(3)& DA3 & W1 \\
10420021$^\dag$ & J19492+4734  &16.2(5)&7.8(3)& DA3 & W3 \\
11337598        & J18577+4909  &22.8(5)&8.6(3)& DA2 & N1,W3 \\
11514682$^\dag$ & J19412+4925  &32.2(5)&7.5(3)& DA1 & W1 \\
11604781        & J19141+4936  & 9.1(5)&8.3(3)& DA5 & W2 \\
11822535$^\dag$ & WDJ1943+500  &36.0(5)&7.9(3)& DA1 & N2\\
\hline
{\tt\ }6862653$^\dag$ & J19267+4219  &\multicolumn{2}{c}{Poor fit}& DB & W2 \\
\hline
\end{tabular}
\end{table}

One of the five DAs, \kic\,11337598, appears to be an unusually
rapid rotator (Fig.~\ref{fig:K11337598}).
A rotational velocity of $v\sin i$\,= 1500\,km\,s$^{-1}$ had to be imposed
on the model in order to get a reasonable fit. This very substantial velocity
corresponds to a rotation period of about 40\,s for a typical WD,
which is about half of the breakup velocity (higher if seen at low
inclination angle). While the spin period is higher than the
\kep~SC sampling frequency, if a modulation with this period were present
in the light curve, it might still show up in the Fourier transform, as the high
sensitivity of \kep\ photometry would have permitted us to see such
high frequencies as reflections across the Nyquist frequency.
Although the fit shown in Fig.~\ref{fig:K11337598} looks
reasonable considering the noise, we cannot completely rule out
a weak magnetic field as a possible alternative, or contributing
effect to the Balmer line broadening. A much higher S/N spectrum
would be required to clearly distinguish these possibilities.
The Kepler light curve shows no high-frequency signal, but does
show a low-amplitude long-period peak (see section~\ref{sect:bins}).

In three cases the spectra do not provide acceptable fits.
When fitting \kic\,7129927, the solution converges to about
\teff\,=\,23\,000\,K and \logg\,=\,7.3, but the cores are clearly
not well fitted (Fig.~\ref{fig:K7129927}).
A new spectrum was recently obtained, and it was
confirmed that the strange features in the cores of the DA2 spectrum
were real.  We conclude that these features can be
explained by a composite DA+DA binary, with the fainter component
having somewhat narrower Balmer lines than the hotter component.
There may be a small shift between the cores of the broad and narrow
components, but this is hard to quantify from the low S/N spectra
currently available. The low S/N also prevents a reliable
decomposition, so deeper spectroscopy of this object is encouraged.

\begin{table*}
\caption[]{Properties of the sdB stars with no significant pulsations.}
\label{tbl:nonpuls}\centering
\begin{tabular}{rlrrrrrrrrrcccl} \hline
\multicolumn{2}{c}{}    &
\multicolumn{3}{c}{100\,--\,500\,\uHz} &
\multicolumn{3}{c}{500\,--\,2000\,\uHz} &
\multicolumn{3}{c}{2000\,--\,8488\,\uHz} &
\multicolumn{4}{c}{Spectroscopic data} \\
         &              &\tsig&\tAmax&\tFmax &\tsig &\tAmax&\tFmax
                        &\tsig&\tAmax&\tFmax &\teff &\logg &\logy   & \\
\tabkic  & Name         &ppm&\tsig&\uHz&ppm&\tsig&\uHz&ppm&\tsig&\uHz&(kK)&(dex)&(dex)& Run\\ \hline
 2020175&J19308+3728& 30&3.4& 482& 29&3.5&1345& 29&3.7&4432&33.0(9)&5.90(5)&--1.5(1)&N1\\
 2304943&J19275+3738& 53&4.8& 365& 50&3.2& 655& 51&3.8&2290&31.2(5)&5.82(7)&--1.7(1)&N2\\
 3353239&J19367+3825& 20&3.0& 497& 20&3.3& 835& 19&3.4&4658&32.4(2)&5.75(5)&--2.7(2)&W1\\
 3527028&J19024+3840& 96&2.9& 376& 93&3.5&1899& 94&3.7&3116&30.1(3)&5.58(5)&--2.7(4)&W2\\
 5340370&J18535+4035& 96&3.0& 107& 94&3.3&1832& 94&3.5&4244&30.2(2)&5.61(4)&--2.4(1)&W2\\
 5938349&J18521+4115& 54&2.9& 457& 54&3.8&1422& 54&4.4&6517&31.9(5)&5.83(6)&--2.6(2)&W1\\
 6522967&J19279+4159& 65&3.5& 217& 63&3.4&1228& 63&4.4&8082&34.3(6)&5.27(9)&--2.7(4)&W1\\
 6614501&J19368+4201& 28&5.3& 365& 27&3.4&1179& 27&4.1&4381&23.1(4)&5.50(5)&--3.0(1)&W1\\
 7104168&FBS1907+425& 18&3.1& 448& 18&3.6& 732& 18&3.8&3557&36.5(5)&5.67(10)&--0.7(1)&N1\\
 7799884&J18456+4335& 57&3.1& 368& 56&4.0& 564& 56&3.6&7327&31.8(4)&5.68(6)&--2.0(1)&N1\\
 8874184&J19084+4508& 50&3.1& 316& 46&4.1&1813& 46&3.8&8361&32.4(9)&5.84(6)&--1.8(1)&N2\\
 9095594&J19369+4526& 74&3.3& 297& 71&3.4&1254& 71&3.7&8010&29.3(4)&5.19(6)&--3.0(1)&W1\\
 9211123&J19144+4539& 28&3.0& 283& 27&3.4&1658& 26&3.8&2033&34.7(4)&5.11(6)&--2.9(1)&N2\\
10149211&J19393+4708& 22&6.9& 370& 22&3.4&1434& 21&3.6&6124&27.6(4)&5.42(5)&--2.7(1)&W1\\
10462707&J19144+4737& 57&3.6& 368& 57&3.9& 561& 57&4.1&8206&28.6(4)&5.25(6)&--3.0(1)&W1\\
10784623&J19045+4810&\multicolumn{9}{c}{Not observed yet}  &29.4(5)&5.44(8)&--2.9(1)&N2\\
10789011&J19136+4808& 23&3.9& 364& 22&3.2&1513& 22&3.9&6243&34.1(2)&5.69(5)&--1.4(1)&N2\\
10961070&J18534+4827&108&3.1& 421&107&3.7& 587&109&4.1&4623&37.4(3)&6.05(5)&--1.0(1)&W2\\
11350152&J19268+4908& 24&3.6& 253& 20&4.2& 630& 19&3.6&5958&35.6(3)&5.57(5)&--1.7(1)&N2\\
11400959&J19232+4917& 51&3.4& 477& 49&3.4& 688& 49&3.9&5568&39.5(4)&6.12(4)&--2.9(1)&W2\\
12021724&J19442+5029& 30&3.3& 436& 29&3.5& 531& 29&3.7&7527&26.2(3)&5.40(4)&--2.3(1)&W1\\
\hline
\end{tabular}\\
{\em Notes.}\,---\,\sig~is the mean of the amplitude spectrum in the region stated.
\Amax~and \Fmax~give the amplitude and frequency of the highest peak.
\end{table*}

The two cases marked with `Poor fit' in Table~\ref{tbl:wds}
are of faint targets obtained in bright sky conditions, and the
background subtraction appears to be inadequate.
The intention when the spectra were made was only
to provide a spectroscopic class for all the targets in the
sample, and in these cases the quality turned out to be too
poor to obtain reliable physical parameters. However, there
is no doubt about the classification. The star listed as a DA
white dwarf shows the broad and deep Balmer lines typical
for DAs not too far from the ZZ\,Ceti instability strip.
The DB is most likely around 16\,000\,K, much too cool to be
a V777\,Her pulsator. At \mkep\,=\,18.2 it is also the
faintest star in our sample.
As none of these stars show any sign of variability above the 4-sigma limit,
we have not attempted
to obtain higher quality spectra.

\subsection{The hot subdwarf stars}

The majority of the stars constituting the current half of our survey
sample are normal sdB or sdOB stars (26 objects), out of which
one is clearly composite with an F/G type companion.
Of the remaining subdwarfs, one is a helium poor sdO star, and
six are He-rich sdO or sdOB stars.
Note that we distinguish between the common
He-sdOB stars that show He\,{\sc i} and He\,{\sc ii} lines with
almost equal depth, and the hotter and rarer He-sdO stars that show
predominantly He\,{\sc ii} lines. The former are seen in various
surveys to form a narrow band at around 40\,000\,K
\citep[e.g.][]{stroer07}, while the latter
do not cluster in the \teff/\logg-plane \citep{ostensen09}.
As in \papI, we have fitted the spectra of the sdB and sdOB stars
to model grids, in order to determine
effective temperature (\teff), surface gravity (\logg),
and photospheric helium abundance
(\logy\,= log\,$N_{\mathrm He}/N_{\mathrm H}$).
The fitting procedure used was the same
as that of \citet{edelmann03}, using the metal-line blanketed
LTE models of solar composition described in \citet{heber00}.
The usual caution about systematic effects, when comparing 
parameters derived from fitting upon grids created using different
methodologies, obviously applies.
For the non-pulsators we list the physical parameters
in Table~\ref{tbl:nonpuls} together with the variability limits
from the frequency analysis discussed in the next section.
The parameters of the pulsators are given in Table~\ref{tbl:pulsators},
together with their variability data.
For the He-rich subdwarfs and the hot sdO we do not provide
physical parameters, as they are beyond the range of our LTE grid.

\begin{table*}
\caption[]{Properties of the non-sdB stars with no significant pulsations.}
\label{tbl:nonsdb}\centering
\begin{tabular}{rlrrrrrrrrrl} \hline
\multicolumn{2}{c}{}    &
\multicolumn{3}{c}{100\,--\,500\,\uHz} &
\multicolumn{3}{c}{500\,--\,2000\,\uHz} &
\multicolumn{3}{c}{2000\,--\,8488\,\uHz} & \\
         &              &\tsig&\tAmax&\tFmax &\tsig &\tAmax&\tFmax
                        &\tsig &\tAmax&\tFmax & Spectroscopic \\
\tabkic  & Survey name  &(ppm)&\tpsig&\tpuHz&(ppm)&\tpsig&\tpuHz
                        &(ppm)&\tpsig&\tpuHz& classification \\ \hline
 2303576 & J19263+3738  &102&2.9& 345&101&3.3& 788&100&3.9&2160& He-sdO \\
 3343613 & J19272+3827  & 43&3.9& 364& 40&3.4&1441& 39&3.7&7517& He-sdOB \\
 5769827 & J18547+4105  & 59&3.1& 236& 56&3.5& 577& 55&3.6&6507& DA0 \\
 6878288 & J19436+4220  & 42&3.6& 148& 42&3.4&1664& 42&3.9&4311& He-sdOB \\
 7129927 & J19409+4240  & 39&3.2& 418& 38&3.4& 960& 38&3.9&5495& DA+DA \\
 7335517 & J18431+4259  & 31&3.6& 187& 33&3.7& 801& 32&3.9&5533& sdO \\
 8054179 & J19569+4350  & 14&3.4& 236& 12&3.3&1499& 12&3.8&2108& He-sdOB \\
10198116 & J19099+4717  & 32&5.5& 369& 31&4.0&1485& 31&3.8&5541& DA3 \\
10207025 & J19260+4716  & 40&3.4& 107& 35&3.3& 591& 35&4.0&7486& He-sdO \\
10449976 & J18472+4741  & 15&3.1& 459& 16&3.3&1200& 16&4.1&4569& He-sdOB \\
11337598 & J18577+4909  & 67&2.9& 288& 64&3.6& 817& 64&3.7&7858& DA1 (rot)\\
11604781 & J19141+4936  & 59&3.1& 350& 59&3.7&1405& 59&3.9&5677& DA5 \\
\hline
\end{tabular}\\
{\em Notes.}\,---\,\sig~is the mean of the amplitude spectrum in the region stated.
\Amax~and \Fmax~give the amplitude and frequency of the highest peak.
\end{table*}

\section{\kep\ photometry}\label{sect:instr}

The \kep\ photometer operates with an intrinsic exposure cycle consisting
of 6.02-s integrations followed by 0.52-s readouts.
The SC photometry is a sum of 9 such integrations,
and LC photometry are a sum of 270 \citep{gilliland10b}.
The LC cycle produces artefacts in the SC light curve, not
just at the LC frequency, \flc\,=\,566.391\,\uHz, but at all harmonics
of this frequency up to the Nyquist frequency, which is
\fnyq\,=\,15\flc\,=\,0.5\fsc\,=\,8496.356\,\uHz.
In most short-cadence light curves, this artefact comb has its
strongest peak at 9\flc\,=\,5098\,\uHz.
For details on the \kep\ data processing pipeline, see \citet{jenkins10}.

The \kep\ spacecraft performs a roll every quarter, in order to keep
its solar panels and sunshield facing the sun. The mission therefore
naturally breaks down into quarterly cycles, and the two quarters
of data analysed here are referred to as Q3 and Q4. Each quarter
is then split into monthly thirds, and the collected
photometric data are downloaded after each such run.
When this happens, the spacecraft
must change its attitude to point its main antenna at the Earth.
During these events, observations cease and the change in pointing
causes a thermal transient in the spacecraft and its instrument.
Afterwards, the spacecraft takes some days to reach an equilibrium
state, and light curves of many targets show deviations, as shifts
in the focal plane slightly
changes the contamination from nearby objects.
Similar thermal transients are seen after unforeseen events cause the
spacecraft to enter safe mode and switch off its detector electronics,
which then takes some time to warm up after resumption of normal operations.
Pointing tweaks also produce discontinuities in faint objects due
to changes in the contamination from nearby objects, but
during Q3 and Q4 no pointing tweaks were required, leaving fewer
corrections necessary than in the earlier quarters.
All datasets could be corrected by one or two continuous curves,
consisting of a leading exponential decay followed by a polynomial
of no more than third order.

Only four events are significant enough to require corrections to the
Q3 and Q4 light curves.
There is a 1-d gap in Q3.1 data, between MJD\,55113.55 and 55114.34,
caused by a loss of fine pointing control.\footnote{
The times used here are modified Julian dates (MJD\,=\,JD\,--\,2500000.5). 
}
\kep~entered safe mode at the very end of Q3.2, so the event did not
affect those light curves significantly, but Q3.3 light curves are
slightly shorter than intended, after the two days of downtime, and
the first few days of these suffer from a more severe thermal excursion
than usual.
In Q4.1, on MJD\,55205, CCD-module 3 failed. The loss of the module
produced temperature drops within the photometer and telescope
structure \citep{kepler_dr06}, and these affect the light curves
in a minor way, similar to a pointing tweak.
The most significant event happened in the middle of Q4.2, when
the spacecraft entered safe mode for 4 full days between
MJD 55229.35 and 55233.31, the longest downtime of the first mission year.

\subsection{Pulsation limits}

In Table~\ref{tbl:nonpuls} and Table~\ref{tbl:nonsdb} we list the limits
from our Fourier analysis of the \kep\ light curves
where no clear pulsations were found, for the non-sdB
and sdB stars respectively.
As in \papI~we provide, for three different frequency ranges, 
the arithmetic mean (which we consider to be the standard deviation, \sig)
of the amplitude spectrum in each frequency range,
and the amplitude (\Amax) and frequency (\Fmax) of the highest peak.
\Amax\ is given as the ratio of the peak amplitude and the \sig\ level.
Frequencies associated with binary and other types of
long-period variability are discussed in Section~\ref{sect:bins}.

As noted in \papI,
peaks with amplitudes as high as 4.1\sig~are seen in many of the
light curves, and we do not consider these to be significant.
Frequencies associated with binary and other types of
long-period variability are discussed in Section 4.3.

For the binaries with harmonics that have significant amplitudes
above 100\,\uHz, we fitted the main period (and harmonic) with
sines as part of the detrending process. The residual light curves
were then sigma-clipped at 4-\sig~to remove outliers before beginning the
Fourier analysis.
Only for the relatively high amplitude binary, \kic\,7335517,
did this procedure leave any significant residuals. For this case,
the \Amax~listed in Table~\ref{tbl:nonsdb}
ignores the first three harmonics of $f_{\rm orb}$\,=\,84.33\,\uHz.

For the non-sdB stars,
the only marginally significant period (5.5\sig) is seen in \kic\,10198116,
at 369\,\uHz. However, structure between 360 and 370\,\uHz~is seen in many
other stars of a wide range of spectral types and observed in various quarters,
and has now been flagged as most likely instrumental in origin.

The limits for the sdB stars that were not
found to show clear pulsations are given in Table~\ref{tbl:nonpuls}.
Here we see three stars that show significant peaks between 364 and
372\,\uHz, which we consider to be spurious. Excluding this frequency
range would drop \Amax~for the low-frequency range to below 3.3\sig~for all
three stars.
Two stars show peaks at 4.4\sig~in the high-frequency domain,
but these peaks are too low to be significant if the light curve
is split into halves, and we therefore consider the evidence
of pulsations in these stars to be too weak to claim detection.
None of the stars show any other peaks higher than 4\sig~in the high
frequency region, either in the full light curve, or in the individual
halves.

\begin{figure*}
\centering
\includegraphics[width=12cm]{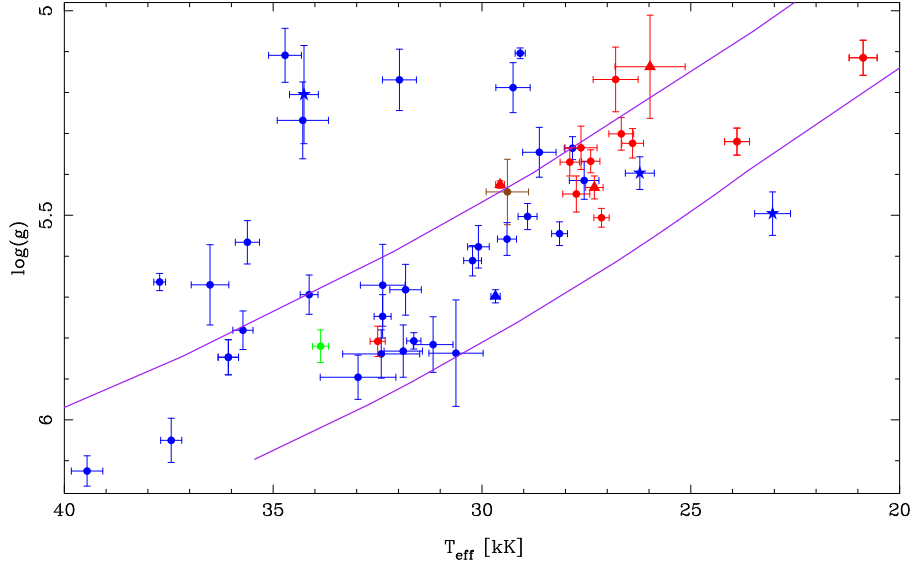}
\caption{The \teff/\logg\ plane for the sdB stars in our sample.
Red symbols indicate the pulsators, blue symbols the non-pulsators, with
the transient pulsator of \papI\ marked with a green bullet and the 
as yet unobserved sdB with a brown symbol.
The apparently single sdBs are marked with bullets, 
the sdB+dM reflection binaries with triangles, 
and the sdB+WDs with stars.
The curves indicate the approximate location of the zero-age and
terminal-age EHB for a canonical sdB model.
}
\label{fig:tgplot}
\end{figure*}

\begin{table*}
\caption[]{Properties of the sdBV stars.}
\label{tbl:pulsators}
\centering
\begin{tabular}{rlcrrrrcccl} \hline
\multicolumn{2}{c}{}    &
\multicolumn{5}{c}{\kep\ data} & \multicolumn{4}{c}{Spectroscopic data} \\
\multicolumn{2}{c}{}    &\tsig&$N_f$&\fmed &\fmin &\fmax &
\teff &\multicolumn{1}{c}{\logg}&\logy & \\
\tabkic & Name      &(ppm)&&(\uHz)&(\uHz)&(\uHz)& (kK) &(dex)&(dex)&Run\\\hline
 7668647&FBS1903+432&22& 15 &173.3&115.9&345.8&27.7(3)&5.45(4)&--2.5(1)&N1\\
 8302197&J19310+4413&41&  6 &183.3&126.3&305.8&26.4(3)&5.32(4)&--2.7(1)&W2\\
10001893&J19095+4659&23& 24 &262.9& 77.5&391.4&26.7(3)&5.30(4)&--2.9(1)&N2\\
10553698&J19531+4743&19& 30 &228.1&104.3&492.9&27.6(4)&5.33(5)&--2.9(2)&W1\\
11558725&J19265+4930&17& 36 &260.5& 78.2&390.9&27.4(2)&5.37(3)&--2.8(1)&W1\\
\hline
\end{tabular}
\end{table*}

As mentioned above, the LC artefacts introduced at $n$\flc\ up to
the Nyquist frequency makes us effectively blind to pulsators with
periods at these frequencies. However, due to the 30\,d length of
the runs, the resolution is sufficiently high that these blind spots
are quite insignificant. A second cause of concern is the Nyquist
limit itself, which at 120\,s represents a period typically seen
in sdB stars. Of the 49 sdBV stars listed in \citet{sdbnot}, 13
have periods at or shorter than 120\,s. The shortest period
reported to date is 78\,s in EC\,01541--1409 \citep{kilkenny09}.
The sdO pulsator, J16007+0748 \citep{woudt06}, is a particular
case of concern since all the 13 periods detected in this star
lie between 57 and 119\,s \citep{rodriguez10a}. However, periods shorter
than the Nyquist limit are still detectable, with a smearing
penalty factor given by  $\sin (x)/x$, where
$x$\,=\,$\pi f \Delta t_{\rm exp}$ \citep{kawaler93}.
The smearing drives all amplitudes to zero as the sampling frequency
is approached, but for most frequencies between \fnyq\ and
the sampling frequency the recovered amplitudes
are still significant. For \kep, we should recover 30 percent
of the amplitude for pulsation periods of 80\,s.
The corresponding frequencies will appear in the FT reflected around
\fnyq, as was seen for the first harmonic of the main pulsation mode
in \kic\,10139564 (Paper~{\sc ii}).

\subsection{The new pulsators}\label{sect:puls}

Five clear V1093\,Her pulsators were detected in 
the second half of the survey phase (Table~\ref{tbl:pulsators}).
The noise level (\sig) was measured in the region 1000\,--\,2500\,\uHz.
In addition to the number of significant pulsation modes
detected in the \kep\ photometry ($N_f$),
and the minimum and maximum pulsation frequencies,
we also provide the power-weighted mean frequency, \fmed. 
The spectroscopic parameters as determined from our fits are also
listed, and all stars are seen to cluster at the hot end of the
$g$-mode region between 26\,000 and 28\,000\,K.

Of the 5 V1093\,Her pulsators described in Paper~{\sc iii},
3 were found to show low-amplitude short-period pulsations
in the frequency range typical for V361\,Hya stars.
Recently, \citet{charpinet10} have concluded that the single short
period peak found in \kic\,2697388 is consistent with a predicted p-mode
in their model that fits the observed g-mode spectrum, thereby making the
case that these objects represent a new kind of hybrid sdBV.

In the current sample, \kic\,7668647 shows two peaks at  4738 and 4739\,\uHz,
at around 5\sig. 
The peaks are most significant in the first half of the run, and
drop below the 4-\sig~limit in the second half of the run.
\kic\,8302197 shows no significant peaks (higher than 4.1\sig)
in the short-period pulsation range.
\kic\,10001893 shows a single 5.7-\sig\ peak at 2925.8\,\uHz.
\kic\,10553698 shows a pair of 4.5-\sig\ peaks at 3073\,\uHz, and
also some structure at 4070\,\uHz.
\kic\,11558725 also shows peaks at 3073\,\uHz, the highest at 8.7\sig.
It is suspicious that two stars show the same frequency, but the
former was observed in Q4.1 and the latter in Q3.3. Thus, it is
not obvious that these are artefacts, and none of the frequencies
found in these stars have been associated with artefacts before.
Thus, it appears that four of the five long-period pulsators might
be hybrid pulsators, but the amplitudes are low so the hybrid nature
needs to be confirmed. See Paper~{\sc vii} for more details.

Fig.~\ref{fig:tgplot} is identical to Fig.~3~in \papI, but with
the new pulsators and non-pulsators added.
Unlike in the first half of the sample, we here clearly have
non-pulsators at \teff\ lower than the transition region
between the $p$- and $g$-mode pulsators (at $\sim$28\,000\,K).

\kic\,6614501 at \teff\,=\,23\,100\,K is unusual as it lies well below
the extreme horizontal branch (EHB) in Fig.~\ref{fig:tgplot}.
It also shows a light curve signature that we interpret as binary,
as discussed in Section~\ref{sect:bins}, below.
This could indicate that this sdB is another example of the rare
post-RGB white dwarf progenitors that are evolving directly from
an RGB evolution interrupted by a common envelope ejection and
towards the white dwarf cooling curve, such as HD\,188112 \citep{heber03}.
Unlike the EHB stars, for which the core managed to reach sufficient mass
for helium ignition before the envelope was ejected, these stars are
much less massive and will become low mass He-core WDs. The current
temperature and gravity of \kic\,6614501 is consistent with the
evolutionary tracks of \citet{driebe99} for a mass of the primary of
$\sim$0.24\,\msol.
Its FT does show some weak peaks around 365\,\uHz, but as mentioned
earlier, these are most likely artefacts. 

\kic\,12021724 is located at 26\,200\,K and has absolutely no significant
peaks in the FT above 35\,\uHz. However, like \kic\,6614501, it shows
a likely binary period. 

\begin{table}
\caption{Binaries and other long-period variables.}
\label{tbl:bins}
\begin{tabular}{rllll} \hline
         &Period& Amp.& Class & Main \\
\tabkic  & (d)  & (\%)&       & variability \\ \hline
 2303576 &0.19206& 0.8 &He-sdO+?& bin/cont\\ 
 3527028 &2.10540& 0.5 & sdB+?  & unknown \\   
 5340370 &0.20--0.72& ...& sdB+?& unknown \\   
 6614501 &0.15746& 0.12& sdB+WD?& binary \\    
 6878288 &3.04065& 1.0 & He-sdOB+? & unknown \\      
 8874184 &2.63670& 0.8 & sdB+?  & var.comp \\  
 7335517 &0.13725& 6.0 & sdO+dM & reflection \\
 8054179 &...    & 0.02& He-sdOB& aperiodic\\  
10149211 &0.60513& 0.5 & sdB+?  & var.comp \\  
10462707 &0.78880& 0.08& sdB+WD?& binary \\    
11337598 &0.09326& 0.04& DA1 (rot) & spin period?\\
11350152 &3.3    & 1.5 & sdB+F/G& var.comp\\   
11604781 &4.87988& 0.5 & DA5    & unknown \\   
12021724 &0.67490& 0.07& sdB+WD?& binary \\    
\hline
\end{tabular}
\end{table}

\subsection{Binaries and other long-period variables}\label{sect:bins}

In Table~\ref{tbl:bins} we list the 14 binaries and
other long-period variables that we have detected in the current
half of the sample. The periods range from 3.3 hours to almost
five days, and the peak-to-peak amplitudes range from a few per cent
to a few hundred ppm.

\begin{figure*}
\centering
\includegraphics[width=13.5cm]{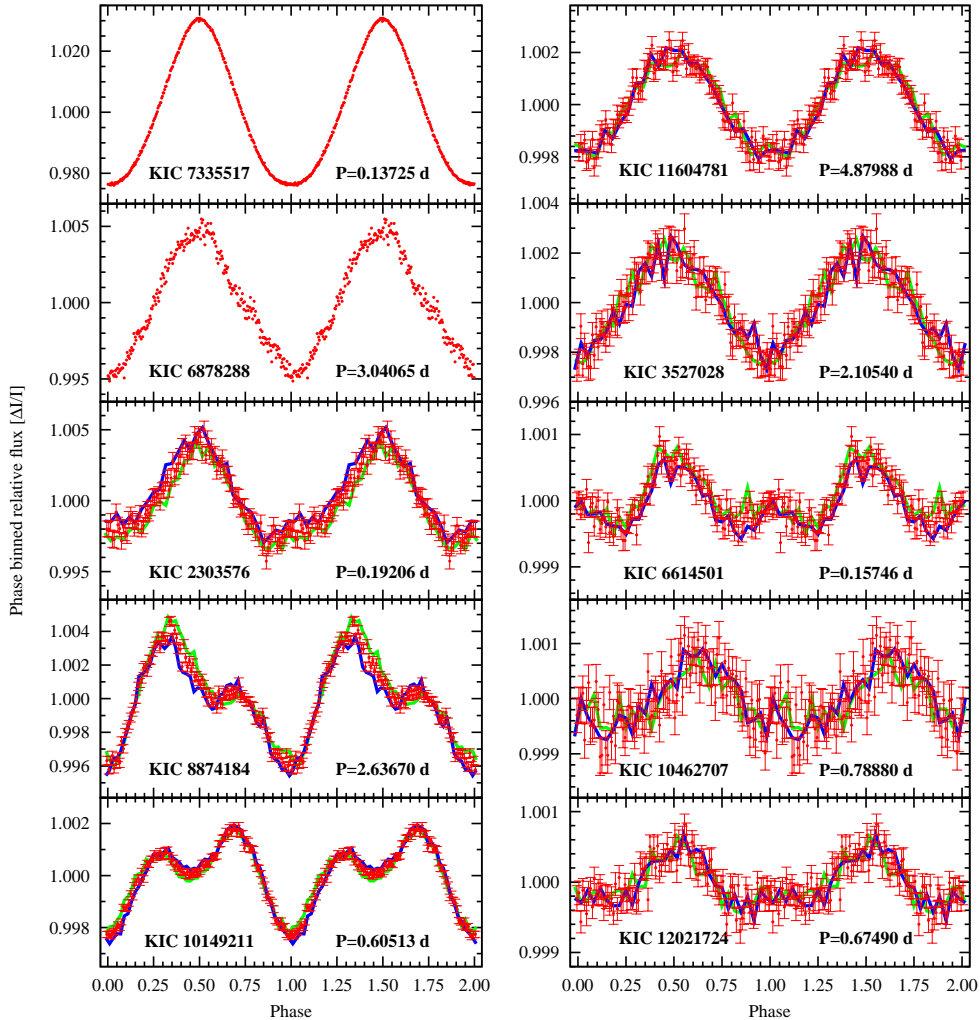}
\caption{Data of ten binary candidates folded on the main
period.
The light curves were folded on the periods given in
Table~\ref{tbl:bins} and repeated on each panel.
For the first two we used 250 bins, and the data are plotted as points.
For the rest we used either 100 bins or 30 bins, due to the lower signal,
and the error bars shown are the rms values for the points in each bin.
We also subdivided the 8 last observation runs into two halves, and
folded these separately. The resulting curves are shown as continuous
green and blue lines.
Two cycles are shown for clarity.}
\label{fig:bins}
\end{figure*}

The sdO star \kic\,7335517 is the clearest binary candidate,
with the maxima slightly sharper than the minima
(Fig.~\ref{fig:bins}, upper left panel), as is typical
for the temperature effects that are seen when a hot subdwarf
irradiates one hemisphere of a cool companion.
The folded light curve is similar to that of KBS\,13, observed
in Q1, and the semi-amplitude is about the same.
We judge the temperature of the primary to be above 40\,000\,K,
so the reflection effect should be much higher than what is
seen, if the system is seen at high inclination.
Thus, the system is most likely seen at low inclination,
as for KBS\,13, or else the companion is substellar.

The remaining 9 objects shown in Fig.~\ref{fig:bins} all display very
low level photometric modulations that may be due to orbital effects,
but we do not consider any of these as being likely to have M-dwarf
companions. A more likely companion would be a white dwarf, in which
case the light curve should display a combination of ellipsoidal deformation and
Doppler beaming, such as seen for KPD\,1946+4340 \citep{bloemen10}.
In particular, \kic\,6614501 has the double-peaked
structure, with alternating maxima and minima of roughly
equal depth, that characterises such beaming binaries.
Splitting the light curve in halves reveals a consistent shape,
as expected for an orbital effect. \kic\,10462707 and \kic\,12021724
are also likely to be sdB+WD binaries, and a WD companion is a possible
interpretation for the He-sdO \kic\,2303576 as well.

\kic\,6878288, 11604781 and 3527028, all show monoperiodic light
curve variations that range between two and five days, and does
not change between the first and second halves of the run.
The He-sdOB \kic\,6878288 and the regular sdB \kic\,3527028
are unlikely to show such long periods intrinsically. But both have 
contamination factors, \corr, indicating that around half the light comes
from other sources near the intended target. This makes it rather futile
to speculate about whether the modulation comes from the the subdwarfs,
close companions, or nearby objects.
\kic\,11604781 is the only regular white dwarf star in the current
sample that shows any long-period modulations.
With \corr\,=\,0.006, any reasonable effect from a contaminating object
is effectively ruled out.
The spectrum is also void of any features indicating a cool
companion, even in the region around H$\alpha$.
For such a cool DA, the companion would have to be substellar
not to contribute to the optical spectrum.
There are no features in our classification spectrum that
can offer clues to the origin of the photometric variability,
such as trace of a companion or magnetic field, so we have
at present no theory that could explain the 4.88\,d signal. 

\kic\,10149211 and 8874184 show variations with a main
period and a strong first harmonic, and small but significant
changes in the shape of the modulation between the first and the second
halves.
Such variations are unlikely to originate from the hot subdwarf
star itself, and with \corr\ between 9 and 24\%\ they
are most likely from a contaminating object.
A possible interpretation is that the contaminating star is heavily 
spotted, and has more spots on one hemisphere than on the other
\citep[see e.g.][]{siwak10}.

\begin{figure}
\centering
\includegraphics[width=\hsize]{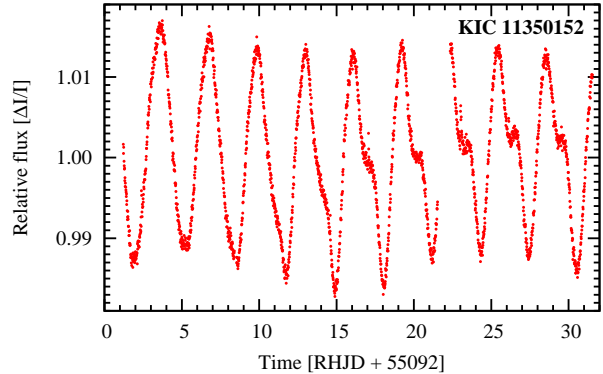}
\caption{The light curve of \kic\,11350152 is dominated by
a period of 3.16\,d, but the shape of the light curve is
changing from cycle to cycle.
The light curve was binned so that each point represents
20 SC exposures, in order to reduce the noise.
}
\label{fig:kic11350152}
\end{figure}

The FT of \kic\,11350152 is dominated by a strong 3.16-d period, and
its first harmonic. There are no other significant peaks in the FT.
As shown in Fig.~\ref{fig:kic11350152}, there are substantial
variations from cycle to cycle.
Also here the most likely interpretation is that the main period seen
is the rotation period of a spotted companion, but a pulsating companion
can not be ruled out. Our blue spectrum shows
clear signatures of an F--G companion, with K and H lines too broad to be
interstellar, and a clear g-band.
The 2MASS IR photometry 
indicates
a rising IR flux, and the object appears single in images, so the
the cool star is likely to be the accretor responsible for stripping
the envelope off the sdB progenitor when it was on the red giant branch.
Also, \corr\ is insignificant so the variations can not
be ascribed to any nearby objects.

\begin{figure}
\centering
\includegraphics[width=8cm]{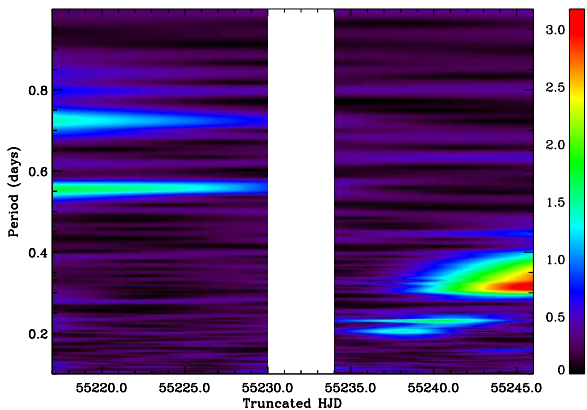}
\caption{
The wavelet transform of \kic\,5340370 shows clear periodicities
at about 0.55 and 0.72\,d in the first part of the light curve.
These apparently disappear completely and are replaced briefly by peaks
at 0.2 and 0.23\,d, which are quickly replaced by a strong and
broad structure between 0.31 and 0.39\,d. The gap
corresponds to the 4\,d safing event in Q4.2.
}
\label{fig:kic05340370}
\end{figure}

\kic\,5340370 shows unusual behaviour below 100\,\uHz.
The periods appear to change completely
between the first half of the run and the second, so we have computed
a wavelet transform (WFT) rather than the regular Fourier transform, 
using the WWZ algorithm of \citet{foster96}. 
The WFT is shown in Fig. \ref{fig:kic05340370}, and in the first half
two clear peaks are found at 0.55 and 0.73\,d (16 and 20\,\uHz).
In the second half, these peaks are barely detectable,
but have been replaced by broad features in the region between between
0.2 and 0.4\,d ($\sim$30 and 60\,\uHz).


\kic\,8054179 shows a light curve with considerable power in the
FT at frequencies below 100\,\uHz, but no clear peaks.
The amplitude of the variability is only at the 200\,ppm level, and
can easily be caused by a contaminating star.
However, it is interesting to note that in the first half of the survey, 
we also saw aperiodic variability in another He-sdOB star, \kic\,9408967,
but then at a somewhat higher amplitude than here.


The WD which we suspected to be an extremely rapid rotator in
section~\ref{sect:wds}, \kic\,11337598, shows no significant peaks
in the FT, except a single low amplitude peak at 124\,\uHz,
(=\,0.093\,d).
With an amplitude of only 360\,ppm this is far too low to be
a binary signal for such a short period orbit, unless the system
is seen improbably close to pole on.
If the 0.093\,d period is instead the spin period of the WD,
then the spectroscopic broadening cannot be caused by
rotation.  A possible explanation could involve
a relatively weak magnetic field causing Zeeman
splitting of the Balmer lines, being sufficiently strong to produce
the observed broadening without producing resolved splitting at
our low S/N. The 0.093\,d photometric period could then be
the spin period of the WD, made visible by weak spots on the
surface of the WD, as observed in WD\,1953--011 \citep{brinkworth05}.
A similar period of 0.0803\,d was observed in GD\,356, although
at ten times the amplitude, and interpreted in a similar way by
\citet{brinkworth04}. With \corr\,=\,0.225 the
signal can also be from a contaminating object,
so no firm conclusions can be made. But better
spectroscopy could easily be invoked to detect any Zeeman splitting
of the Balmer lines, or confirm our original hypothesis of rapid
rotation. In either case, \kic\,11337598 is an intriguing WD.

\section{Discussion and conclusions}

We have completed a survey for compact pulsators with \kep, and
as far as the subdwarf B stars are concerned the survey has
been a great success. Unfortunately, no pulsating white dwarfs
were found in the survey sample. This is not entirely surprising
as the number of DA white dwarfs surveyed is only 13, and of those
only a couple are anywhere close to the ZZ\,Ceti instability strip.
One star that sits slightly above the instability region is \kic\,10420021,
observed in Q2.2. As we noted in \papI, it does have a 4.5-\sig~peak
in the FT at 196.4\,\uHz. It was reobserved for three months
in Q5, but a quick analysis of this recently released light curve
reveals no trace of any significant signals at periods shorter than
a few days.

The spectroscopic survey used to describe the targets in \papI\ and
in this paper was conducted mostly after the deadline for submitting targets
for the \kep~survey phase had passed. If such a survey had been
made earlier, it would have been evident that the WD sample was
too small to have a significant chance of containing WD pulsators.
But if more stars fainter than \mkep\,=\,17.5 had been retained in the sample,
the chance of finding WD pulsators would have been much higher,
since at such faint magnitudes hot subdwarfs will have to be located
beyond the Galactic disk, and no longer dominate UV-selected samples.
As we have seen from the few stars in our sample that
have magnitudes close to \mkep\,=\,18, the \kep~photometry is
still excellent with 4-sigma detection limits around the millimagnitude
level, even after taking into account substantial contamination factors.

That the survey did not reveal any sdO pulsators was not
a surprise, since only one such object have been revealed to date,
implying that these objects are exceedingly rare. Of the two objects
in the survey classified as sdO+F/G binaries, similar to the prototype
J16007+0748 \citep{woudt06}, one (\kic\,9822180) did show a marginal
peak and will be reobserved for three months in Q6.
In the current half of the sample
only one sdO star is not helium rich, and that object, \kic\,7335517,
is the clearest photometric binary in the sample. This object is
most likely much cooler than J16007+0748, and more similar to the 
eclipsing sdO+dM
binary AA\,Dor, but seen at a low inclination angle.

During the first year of the \kepmi, 
we have surveyed 32 sdB pulsator candidates hotter than 28\,000\,K,
and found only one clear and unambiguous V361\,Hya pulsator
(Paper~{\sc ii}).
One other sdB star shows a single significant short-period frequency that
drops systematically in amplitude until it is below the detection
limit (\papI).
A third pulsator, 2M1938+4603, was found in an eclipsing binary,
and shows an exceptionally rich pulsation spectrum, that includes short,
long and intermediate periods. However, it has no strong (above 500\,ppm)
pulsation modes, which makes it quite an exceptional hybrid pulsator.
All these stars were
discovered in the first half of the survey sample, and the
current sample of 17 stars above 28\,000\,K
contains no further V361\,Hya pulsators.
This means that the number of short-period pulsators found in
the \kep\ sample is actually less than the 10 percent fraction that has been
found in ground-based surveys \citep{sdbnot},
at least when considering that 2M1938+4603 would have been almost
impossible to recognise as a pulsator from the ground.

For the V1093\,Her pulsators, in the first half of the survey we found only
one star below 28\,000\,K that was not pulsating, and that one was less
than one \sig~below this temperature limit. That led us to the preliminary
conclusion that all sdBs below 28\,000\,K may be pulsators.
However, in the current sample we find that only 5 out of 8
stars below 28\,000\,K are pulsators. Only one of these are
within 1-\sig~of the temperature boundary. The remaining two appear to
be sdB+WD short-period binaries, judging from the long-period variations
in their light curves.
These are also the only two short-period sdB+WD binaries
we have been able to identify in the current sample. We do not
see any reason why pulsations should be systematically suppressed
in sdB+WD binaries. After all, the well-studied V361\,Hya star
KL\,UMa is known to be in a $P$\,=\,0.376\,d binary with a WD
companion \citep{otoole04}, and V2214\,Cyg in a $P$\,=\,0.095\,d
binary \citep{geier07}. However, as pointed out by
\citet{ostensen09}, KL\,UMa is the only well-studied
sdBV of those known to be located in the boundary region
between the V361\,Hya and the V1093\,Her pulsators for which
hybrid pulsations has not been detected.
One may speculate that g-modes can be suppressed somehow in
sdB+WD binaries, but this is not the case. Of the 9 V1093\,Her stars
described in the literature, 3 have been published as sdB+WD
binaries; HZ\,Cnc \citep[$P$\,=\,27.81\,d, ][]{morales-rueda03},
V2579\,Oph \citep[$P$\,=\,0.83\,d, ][]{for06} and
PG\,0101+039 \citep[$P$\,=\,0.57\,d, ][]{geier08}.
However, in the case of \kic\,6614501, its position below the canonical
EHB is unexpected for an sdB that has evolved through common-envelope
ejection. The absence of pulsations in this star can therefore
be explained if it is a low-mass post-RGB star that has not ignited
helium in its core rather than a regular EHB star, as such stars
would evolve too rapidly to build up a Z-bump that can drive
pulsations.

The total fraction of pulsators below 28\,000\,K ended up to be
12 out of 16, or 75 percent, the same number as the rough estimate
given by \citet{green03}. We are still puzzled by the large fraction
of sdB stars that show no trace of pulsations in spite of the unprecedented
duration and low noise level provided by \kep. We had expected that the
fraction of pulsators would increase with increasing precision, but
evidently this has not happened.

Thanks to the exceptional precision of the \kep\ measurements, we can
now conclude that there certainly are sdB stars, both on the hot and
cold end of the EHB, that show no trace of pulsations. 
Possible explanations for the non-pulsators would have to answer
why the pulsation driving mechanism is suppressed in some EHB stars and not
others.
They may represent a low-metalicity population that either started out 
being low in iron-group elements, or evolved to have low metalicity envelopes,
as may be the case in merger models. Time dependent changes in the iron
profiles as discussed by \citet{fontaine06a, fontaine06b} can also explain
non-pulsators; 
They may be the youngest among the EHB population, for which an iron opacity
bump in the driving region has yet to accumulate, or they can be the
oldest EHB stars for which low-level winds have depleted the iron reservoir.

Another significant result of our survey is that
many of the V1093\,Her pulsators show
signs of hybrid behaviour, with single low-amplitude modes in the
high frequency region. The V361\,Hya pulsator we found also displays
a single mode in the long-period region. This indicates that hybrid
behaviour is not unusual for sdBV stars regardless of their position
on the EHB, and not confined to the DW\,Lyn stars that sits on
the boundary between the short- and long-period pulsators in the
\teff/\logg~plane. With \kep\ targeting these pulsators at
regular intervals throughout its mission, we will soon know if these
low level hybrid modes are transient or persistent features of the
pulsation spectra of these stars.

\section*{Acknowledgments}

The authors gratefully acknowledge the \kep\ team and all who
have contributed to enabling the mission.
Funding for the \kepmi\ is provided by NASA's Science Mission Directorate.

The research leading to these results has received funding from the European
Research Council under the European Community's Seventh Framework Programme
(FP7/2007--2013)/ERC grant agreement N$^{\underline{\mathrm o}}$\,227224
({\sc prosperity}), as well as from the Research Council of K.U.Leuven grant
agreement GOA/2008/04.
AB gratefully appreciates funding from Polish Ministry of Science and Higher
Education under project N$^{\underline{\mathrm o}}$\,554/MOB/2009/0.
SC thanks the Programme National de
Physique Stellaire (PNPS, CNRS/INSU, France) for financial support.
ACQ is supported by the Missouri Space Grant funded by NASA.

For the spectroscopic observations presented here we acknowledge the Nordic
Optical Telescope at the Observatorio del Roque de los Muchachos (ORM)
on La Palma, operated jointly by Denmark, Finland, Iceland,
Norway, and Sweden, and the William Herschel and Isaac Newton telescopes
also at ORM, operated by the Isaac Newton Group.

\bibliographystyle{mn2e}
\bibliography{sdbrefs}

\label{lastpage}

\end{document}